\documentclass[aps,prd,twocolumn,superscriptaddress,floatfix]{revtex4}
\usepackage{bm}
\usepackage[dvipdfmx]{graphicx}
\usepackage{amsmath,amssymb,times}
\usepackage{ascmac}
\usepackage{braket}

\numberwithin{equation}{section}

\newcommand{\bequ}{\begin{equation}}
\newcommand{\eequ}{\end{equation}}
\newcommand{\bea}{\begin{eqnarray}}
\newcommand{\eea}{\end{eqnarray}}

\DeclareSymbolFont{boldletters}{OML}{cmm} {b}{it}
\DeclareSymbolFontAlphabet{\mathbit}{boldletters}
\DeclareMathSymbol{\alpha}{\mathalpha}{letters}{"0B}
\DeclareMathSymbol{\beta}{\mathalpha}{letters}{"0C}
\DeclareMathSymbol{\gamma}{\mathalpha}{letters}{"0D}
\DeclareMathSymbol{\delta}{\mathalpha}{letters}{"0E}
\DeclareMathSymbol{\epsilon}{\mathalpha}{letters}{"0F}
\DeclareMathSymbol{\zeta}{\mathalpha}{letters}{"10}
\DeclareMathSymbol{\eta}{\mathalpha}{letters}{"11}
\DeclareMathSymbol{\theta}{\mathalpha}{letters}{"12}
\DeclareMathSymbol{\iota}{\mathalpha}{letters}{"13}
\DeclareMathSymbol{\kappa}{\mathalpha}{letters}{"14}
\DeclareMathSymbol{\lambda}{\mathalpha}{letters}{"15}
\DeclareMathSymbol{\mu}{\mathalpha}{letters}{"16}
\DeclareMathSymbol{\nu}{\mathalpha}{letters}{"17}
\DeclareMathSymbol{\xi}{\mathalpha}{letters}{"18}
\DeclareMathSymbol{\pi}{\mathalpha}{letters}{"19}
\DeclareMathSymbol{\rho}{\mathalpha}{letters}{"1A}
\DeclareMathSymbol{\sigma}{\mathalpha}{letters}{"1B}
\DeclareMathSymbol{\tau}{\mathalpha}{letters}{"1C}
\DeclareMathSymbol{\upsilon}{\mathalpha}{letters}{"1D}
\DeclareMathSymbol{\phi}{\mathalpha}{letters}{"1E}
\DeclareMathSymbol{\chi}{\mathalpha}{letters}{"1F}
\DeclareMathSymbol{\psi}{\mathalpha}{letters}{"20}
\DeclareMathSymbol{\omega}{\mathalpha}{letters}{"21}
\DeclareMathSymbol{\varepsilon}{\mathalpha}{letters}{"22}
\DeclareMathSymbol{\vartheta}{\mathalpha}{letters}{"23}
\DeclareMathSymbol{\varpi}{\mathalpha}{letters}{"24}
\DeclareMathSymbol{\varrho}{\mathalpha}{letters}{"25}
\DeclareMathSymbol{\varsigma}{\mathalpha}{letters}{"26}
\DeclareMathSymbol{\varphi}{\mathalpha}{letters}{"27}
\DeclareMathSymbol{\Gamma}{\mathalpha}{letters}{"00}
\DeclareMathSymbol{\Delta}{\mathalpha}{letters}{"01}
\DeclareMathSymbol{\Theta}{\mathalpha}{letters}{"02}
\DeclareMathSymbol{\Lambda}{\mathalpha}{letters}{"03}
\DeclareMathSymbol{\Xi}{\mathalpha}{letters}{"04}
\DeclareMathSymbol{\Pi}{\mathalpha}{letters}{"05}
\DeclareMathSymbol{\Sigma}{\mathalpha}{letters}{"06}
\DeclareMathSymbol{\Upsilon}{\mathalpha}{letters}{"07}
\DeclareMathSymbol{\Phi}{\mathalpha}{letters}{"08}
\DeclareMathSymbol{\Psi}{\mathalpha}{letters}{"09}
\DeclareMathSymbol{\Omega}{\mathalpha}{letters}{"0A}

\def\fun#1#2{\lower3.6pt\vbox{\baselineskip0pt\lineskip.9pt
\ialign{$\mathsurround=0pt#1\hfil##\hfil$\crcr#2\crcr\sim\crcr}}}

\begin{document}
\title{
QCD-inequality analyses on pion condensate at
real and imaginary isospin chemical potentials \\
under finite imaginary quark chemical potential
}

\author{Junpei Sugano}
\email[]{sugano@phys.kyushu-u.ac.jp}
\affiliation{Department of Physics, Graduate School of Sciences, Kyushu University,
             Fukuoka 819-0395, Japan}

\author{Hiroaki Kouno}
\email[]{kounoh@cc.saga-u.ac.jp}
\affiliation{Department of Physics, Saga University,
             Saga 840-8502, Japan}

\author{Masanobu Yahiro}
\email[]{yahiro@phys.kyushu-u.ac.jp}
\affiliation{Department of Physics, Graduate School of Sciences, Kyushu University,
             Fukuoka 819-0395, Japan}

\date{\today}

\begin{abstract}
 By employing QCD inequalities,
 we discuss appearance of the pion condensate
 for both real and imaginary
 isospin chemical potentials.
 In our discussion, imaginary quark chemical potential
 is also taken into account.
 We show that the charged pion can condense
 for real isospin chemical potential,
 but not for imaginary one. 
 Furthermore, we evaluate the expectation value of
 the neutral-pion field $\braket{\pi^3}$
 for imaginary isospin chemical potential
 by using framework of the twisted mass.
 As a result, it is found that
 $\braket{\pi^3}$ becomes zero for the
 finite current-quark mass, whereas
 the expression of $\braket{\pi^3}$ gives
 the Banks-Casher relation in the massless limit.
\end{abstract}

\maketitle

\section{INTRODUCTION}
Mass spectrum of hadronic degree of freedoms
is a key to understand 
properties of the QCD in the low-energy regime.
QCD inequalities provide a powerful
framework to deduce the relation
between various hadron masses,
directly from the QCD~\cite{Weingarten, Witten, Nussinov, Espriu}.
The inequalities are also useful to see
which symmetries are spontaneously
broken or not~\cite{Vafa-Witten}, and
which kind of meson condensates can occur
under various external variables, such as
isospin chemical potential
\cite{Kogut1, Kogut2, Son-Stephanov}.
As a review, for example, see Ref.~\cite{Nussinov_review}.
Application of QCD inequalities is extended to
analyses on free energy of the QCD~\cite{Cohen}, the QCD 
in the large $N_{\rm c}$
limit~\cite{Hidaka-Yamamoto},
and hadron interactions~\cite{Detmold}.

In applying QCD inequalities,
it is necessary that 
the measure in the QCD grand-canonical partition 
function has positivity~\cite{Nussinov_review}. 
This is closely related to
whether the fermion determinant possesses positivity or not.
It is well-known that
the fermion determinant becomes complex
for non-zero quark chemical potential 
($\mu_{q}$)~\cite{Forcrand_sign_problem}.
Meanwhile, positivity is ensured for imaginary $\mu_{q}$
\cite{Roberge-Weiss}
and hence QCD inequalities are applicable there.
The introduction of imaginary $\mu_{q}$ also
plays a critical role in
lattice QCD (LQCD) simulations
\cite{Forcrand-Philipsen, DElia-Lombardo, DElia-Sanfilippo,
Nagata-Nakamura, Bonati,
Bonati-21, Takahashi, Sugano_screening-mass},
since usual Monte-Calro technique can be applied.

At finite isospin chemical potential ($\mu_{I}$)
and imaginary $\mu_{q}$,
positivity is also realized whether $\mu_{I}$ is real
\cite{Alford, Kogut-Sinclair, Kashiwa-isospin}
or imaginary
\cite{Sakai_JPhys}.
LQCD simulations are thus feasible in both the cases.
Indeed, various quantities
were calculated by using LQCD simulations so far
\cite{Kogut-Sinclair, Cea, DElia-imaginary-iso,
Detmold-Orginos-Shi}.
These results may give a hint to understand
behavior of highly isospin asymmetric matter
that exists in the interior of neutron stars~\cite{Shapiro}.

The studies on finite real $\mu_{I}$
are also seen in Refs.~\cite{Kogut1, Kogut2, Son-Stephanov},
based on QCD inequalities and the chiral perturbation theory
in which $\mu_{q}$ is set to zero.
It was proved in Ref.~\cite{Son-Stephanov} that
the charged-pion condensate occurs for real $\mu_{I}$,
which is starting at $\mu_{I}=m_{\pi}/2$ with the pion mass
$m_{\pi} \sim 138$ MeV.
Meanwhile, Sakai \textit{et al.}
studied the imaginary $\mu_{I}$ region
in Ref.~\cite{Sakai_JPhys}
with the chiral perturbation theory and
the Polyakov-loop extended Nambu--Jona-Lasinio model
\cite{Fukushima, Ratti1, Ratti2, Rossner}.
They demonstrated that there is no pion condensate in the
entire region of imaginary $\mu_{I}$.
It is interesting to clarify the reason why the pion
condensate does not take place at imaginary $\mu_{I}$
in the view point of QCD inequalities,
considering the contributions of imaginary $\mu_{q}$ simultaneously.

In this paper, we employ QCD inequalities
to the real or imaginary $\mu_{I}$ regions,
taking also into account imaginary $\mu_{q}$.
We first investigate the $\gamma_{5}$-hermiticity
of the fermion matrix.
It is shown that
positivity of the fermion determinant
is guaranteed for both cases,
but expression of the $\gamma_{5}$-hermiticity is different
between them.
Next, we derive QCD inequalities
and prove that the charged-pion
condensate can take place for real $\mu_{I}$,
whereas there is no charged-pion condensate for imaginary $\mu_{I}$.
This suggests that
the result in Ref.~\cite{Son-Stephanov}
holds even if imaginary $\mu_{q}$ switches on.

As for the neutral pion $\pi^3$,
QCD inequalities are not available
since the $\pi^3$ channel has a disconnected piece in its correlator.
Therefore,
the expectation value $\braket{\pi^3}$ 
of the neutral-pion field 
is evaluated directly.
To do so, we use the twisted-mass technique~\cite{Alpha_Coll, Shindler, Hansen, SV}.
From the analysis, we find that $\braket{\pi^3}$
vanishes for the finite current-quark mass.
Meanwhile, the Banks-Casher relation~\cite{BC} is deduced
in the massless limit.

The rest of this paper is organized as follows. 
In Sec.~\ref{Fermion_Determinant}, we discuss 
the $\gamma_{5}$-hermiticity of the fermion matrix
and positivity of the measure.
In Sec.~\ref{Sec_QCD_inequality},
we formulate QCD inequalities for the pion channel
and study the possibility of appearance of the pion condensate.
In Sec.~\ref{Sec_exp_pi3},
we evaluate the expectation value of the neutral-pion
field. Section~\ref{Sec_Summary} is devoted to a summary.

\section{FERMION DETERMINANT AND $\gamma_{5}$-HERMITICITY}
\label{Fermion_Determinant}
Our starting point is the
two-flavor QCD Lagrangian with finite $\mu_{q}$ 
in Euclidean space-time:
\begin{eqnarray}
 \mathcal{L}_{\rm QCD}
  =\bar{q}(\gamma_{\mu}D_{\mu}+\hat{m}-\mu_{q}\gamma_{4})q
  +\frac{1}{4g^2}F^a_{\mu\nu}F^a_{\mu\nu},
  \label{QCD_Lagrangian_realmu}
\end{eqnarray}
where $q=(u,d)^{\rm T}$ is the quark field,
$D_{\mu}=\partial_{\mu}+iA_{\mu}$ is
the covariant derivative,
and $F^a_{\mu\nu}$ is the field strength of
the gluon field $A_{\mu}$.
The current quark-mass matrix $\hat{m}$
is given by $\hat{m}=\textrm{diag}(m_{u},m_{d})$
with current $u$- and $d$-quark masses.
Here, the condition $m_{u}\ne m_{d}$ is imposed
unless otherwise stated.
In the following discussion,
we do not consider the $\theta$ term
that breaks \textit{CP} symmetry
\cite{Belavin-Polyakov-Schwartz-Tyupkin,tHooft-PRL37,tHooft-PRD14-18}
since it causes the sign problem~\cite{Vafa-Witten, Sasaki_EPNJL},
even if the fermion determinant has positivity.

From Eq.~(\ref{QCD_Lagrangian_realmu}),
we can define the QCD action
and the QCD grand-canonical partition function
as
\begin{align}
 &S_{\rm QCD}=\int^{\beta}_{0}d\tau\int d^3\mathbf{x}\mathcal{L}_{\rm QCD},
 \label{QCD_action}
 \\
 &Z_{\rm QCD}
  =\int\mathcal{D}A_{\mu}\mathcal{D}\bar{q}\mathcal{D}q
  \exp
   \left[
   -S_{\rm QCD}
  \right],
 \label{QCD_partition_function}
\end{align}
where $\beta=1/T$ is an inverse temperature ($T$).
The gluon and quark fields satisfy the 
boundary conditions,
\begin{eqnarray}
 A_{\mu}(\tau+\beta,\mathbf{x})=A_{\mu}(\tau,\mathbf{x}),\ \
  q(\tau+\beta,\mathbf{x})=-q(\tau,\mathbf{x}),
\end{eqnarray}
for Euclidean-time ($\tau$) direction.
In Eq.~(\ref{QCD_partition_function}),
the quark field appears only as a bilinear form
an can be integrated out:
\begin{align}
 &Z_{\rm QCD}=\int\mathcal{D}A_{\mu}
  \textrm{Det}\mathcal{M}(\mu_{q})\textrm{e}^{-S_{\rm G}}
  \equiv \int \mathcal{D}\mu(A),
 \label{measure_QCD}
 \\
 & \mathcal{D}\mu(A)
 =\mathcal{D}A_{\mu}\textrm{Det}\mathcal{M}(\mu_{q})\textrm{e}^{-S_{\rm G}},
 \label{actual_form_measure}
\end{align}
where $S_{\rm G}$ is the pure gauge action
and $\mathcal{M}(\mu_{q})$ is the two-flavor fermion matrix defined by 
\begin{eqnarray}
 \mathcal{M}(\mu_{q})=\gamma_{\mu}D_{\mu}+\hat{m}-\mu_{q}\gamma_{4}.
\end{eqnarray}
The symbol ``Det'' in Eq.~(\ref{actual_form_measure})
stands for the determinant
for flavor, Dirac, and color indices.

For $\mu_{q}=0$,
the fermion determinant $\textrm{Det}\mathcal{M}(\mu_{q})$
and the measure~(\ref{measure_QCD})
has positivity~\cite{Stephanov_QCD-phase-diagram} because
the fermion matrix has the following $\gamma_{5}$-hermiticity
\begin{eqnarray}
 \gamma_{5}\mathcal{M}(0)\gamma_{5}=(\mathcal{M}(0))^{\dag}.
  \label{gamma5_hermiticity}
\end{eqnarray}
For finite real $\mu_{q}$, however, 
the $\gamma_{5}$-hermiticity is lost
due to the relation
\cite{Forcrand_sign_problem}
\begin{eqnarray}
  \gamma_{5}\mathcal{M}(-\mu_{q})\gamma_{5}
  =(\mathcal{M}(\mu_{q}))^{\dag},
  \label{sign_problem}
\end{eqnarray}
which induces the sign problem
and positivity of the measure
(\ref{measure_QCD}) is not ensured any longer.

One of the solutions to recover positivity
is an introduction of imaginary chemical potential
$\mu_{q}=i\theta_{q}T$ with
dimensionless quark chemical potential $\theta_{q}$.
Indeed, the relation
\begin{eqnarray}
 \gamma_{5}\mathcal{M}(i\theta_{q}T)\gamma_{5}
  =(\mathcal{M}(i\theta_{q}T))^{\dag}
\end{eqnarray}
guarantees positivity of the measure
\cite{Stephanov_QCD-phase-diagram}.

Now,
let us consider the case of finite isospin chemical potential, i.e.
$\mu_{\rm I} > 0$ or $\mu_{\rm I} < 0$.
In this case, $\mu_{q}$ and $\mu_{I}$
are given by
\begin{eqnarray}
 \mu_{q}=\frac{\mu_{u}+\mu_{d}}{2},\ \ 
  \mu_{I}=\frac{\mu_{u}-\mu_{d}}{2},
\end{eqnarray}
where $\mu_{u}$ and $\mu_{d}$ are the
$u$- and $d$-quark chemical potentials.
Inversely, $\mu_{u}$ and $\mu_{d}$ are
\begin{eqnarray}
 \mu_{u}=\mu_{q}+\mu_{I},\ \ \mu_{d}=\mu_{q}-\mu_{I},
  \label{muu-mud}
\end{eqnarray}
respectively.
For finite $\mu_{I}$,
the QCD Lagrangian is changed into
\begin{eqnarray}
 \mathcal{\tilde{L}}_{\rm QCD}
  =\mathcal{L}_{\rm QCD}-\mu_{I}\bar{q}\gamma_{4}\tau^3q
  \label{QCD_Lagrangian_iso}
\end{eqnarray}
and the isospin SU(2) symmetry is
explicitly broken to $\textrm{U}(1)_{\textrm{I}_{3}}$,
where $\textrm{I}_{3}=\tau^3/2$
for the third component $\tau^3$ of the Pauli matrix.
The fermion determinant is thus rewritten into
\begin{eqnarray}
 \mathcal{\tilde{M}}(\mu_{q},\mu_{I})
  =\gamma_{\mu}D_{\mu}+\hat{m}
  -\mu_{q}\gamma_{4}-\mu_{I}\gamma_{4}\tau^3.
  \label{fermion_matrix_iso}
\end{eqnarray}

We first consider the case that $\mu_{I}$ is real.
Under the setting of $m_{u}=m_{d}=m_{0}$,
Eq.~(\ref{fermion_matrix_iso}) satisfies
the relation
\begin{align}
 \tau^a\gamma_{5}\mathcal{\tilde{M}}(i\theta_{q}T,\mu_{I})\gamma_{5}\tau^a
  =\mathcal{\tilde{M}}^{\dag}(i\theta_{q}T,\mu_{I})\ \ (a=1,2) ,
  \label{gamma5_hermiticity_iso}
\end{align}
where $\tau^a$ means the first or the second component of the Pauli
matrix.
Here, the summation for $a$ is not taken.
From Eq.~(\ref{gamma5_hermiticity_iso}),
it can be proved that
the fermion determinant $\textrm{Det}\mathcal{\tilde{M}}(i\theta_{q}T,\mu_{I})$
possesses positivity~\cite{Kouno_isospin} because
\begin{align}
 \Bigl\{\textrm{Det}&\mathcal{\tilde{M}}(i\theta_{q}T,\mu_{I})\Bigr\}^{\ast}
 \notag \\
 &=\Bigl\{
 \textrm{det}\mathcal{M}'(i\theta_{q}T+\mu_{I})
 \textrm{det}\mathcal{M}'(i\theta_{q}T-\mu_{I})
 \Bigr\}^{\ast}
 \notag \\
 &=\textrm{det}\mathcal{M}'(i\theta_{q}T-\mu_{I})
 \textrm{det}\mathcal{M}'(i\theta_{q}T+\mu_{I})
 \notag \\
 &=\left|\textrm{det}\mathcal{M}'(i\theta_{q}T+\mu_{I}) \right|^2
 \notag \\
 &=\textrm{Det}\mathcal{\tilde{M}}(i\theta_{q}T,\mu_{I})\ge 0,
 \label{complex-conjugate_imaginary-muq_real-iso}
\end{align}
where
\begin{align}
 \mathcal{M}'(i\theta_{q}T\pm\mu_{I})
 =\gamma_{\mu}D_{\mu}+m_{0}-(i\theta_{q}T\pm\mu_{I})\gamma_{4}
 \label{one-flavor_fermion_matrix}
\end{align}
is the one-flavor fermion matrix with $i\theta_{q}T\pm\mu_{I}$
and the symbol ``det'' denotes the determinant only for
Dirac and color indices.
Note that
\begin{align}
  \gamma_{5}\mathcal{M}'(i\theta_{q}T\pm \mu_{I})\gamma_{5}
 =\mathcal{M}'(i\theta_{q}T\mp \mu_{I})
\end{align}
and hence
\begin{align}
  \left\{\textrm{det}\mathcal{M}'(i\theta_{q}T\pm\mu_{I})\right\}^{\ast}
   =\textrm{det}\mathcal{M}'(i\theta_{q}T\mp\mu_{I}).
\end{align}
The measure
\begin{eqnarray}
 \mathcal{D}\tilde{\mu}(A)
  =\mathcal{D}A_{\mu}\textrm{Det}
  \mathcal{\tilde{M}}(i\theta_{q}T,\mu_{I})\textrm{e}^{-S_{\rm G}}
  \label{measure_iso}
\end{eqnarray}
thus maintains positivity.
Along this line,
we also call Eq.~(\ref{gamma5_hermiticity_iso})
the $\gamma_{5}$-hermiticity.

Now, we return to the condition $m_{u}\ne m_{d}$
and show that
the fermion determinant also keeps positivity
for finite imaginary isospin chemical potential,
i.e. $\mu_{I}=i\theta_{I}T$ with
dimensionless isospin chemical potential $\theta_{I}$.
For $\mu_{I}=i\theta_{I}T$,
the fermion matrix does \textit{not} satisfy
Eq.~(\ref{gamma5_hermiticity_iso}),
but rather fulfills
\begin{eqnarray}
 \gamma_{5}\tilde{\mathcal{M}}(i\theta_{q}T,i\theta_{I}T)\gamma_{5}
  =\mathcal{\tilde{M}}^{\dag}(i\theta_{q}T,i\theta_{I}T).
  \label{gamma5_hermiticity_imaginaryiso}
\end{eqnarray}
It should be noted that the Pauli matrix $\tau^a$
and the condition $m_{u}=m_{d}$ are not needed
to prove Eq.~(\ref{gamma5_hermiticity_imaginaryiso}).

From this, its determinant
\begin{align}
 \textrm{Det}\tilde{\mathcal{M}}(i\theta_{q}T,i\theta_{I}T)
 &=\textrm{det}\mathcal{M}'(i\theta_{u}T)\textrm{det}\mathcal{M}'(i\theta_{d}T)
\end{align}
have positivity, since the relation
\begin{align}
 \begin{split}
  & \gamma_{5}\mathcal{M}'(i\theta_{f}T)\gamma_{5}
  =(\mathcal{M}'(i\theta_{f}T))^{\dag}
 \end{split}
\end{align}
is satisfied for $f=u, d$ and this type of $\gamma_{5}$-hermiticity
guarantees positivity~\cite{Stephanov_QCD-phase-diagram}.
Here, we have used Eq.~(\ref{one-flavor_fermion_matrix})
and introduced $\theta_{u}, \theta_{d}$ as
\begin{eqnarray}
 \theta_{u}=\theta_{q}+\theta_{I},\ \ \theta_{d}=\theta_{q}-\theta_{I}.
\end{eqnarray}
Positivity of the corresponding measure is thus ensured.

{\tabcolsep = 1mm
\begin{table}[t]
    \begin{center}
     \caption{In this table, we present whether positivity exists or not
     for each case. The word ``not'' means that
     positivity of the measure does not exist
     in the corresponding case.}
  \begin{tabular}{c|c|c}
   \hline \hline
   & non-zero real $\mu_{q}$ & imaginary $\mu_{q}$
   \\ \hline
   real $\mu_{I}$ &not & has positivity for $m_{u}=m_{d}$
           \\
   imaginary $\mu_{I}$ & not & has positivity for any $m_{u}$ and $m_{d}$
           \\
   \hline \hline
  \end{tabular}
   \label{table_positivity}
    \end{center}
\end{table}
}

From the discussions 
mentioned above,
we can apply QCD inequalities to
the cases of
imaginary $\mu_{q}$ and real $\mu_{I}$,
or imaginary $\mu_{q}$ and imaginary $\mu_{I}$;
see Table~\ref{table_positivity}.
In the next section,
we formulate QCD inequalities 
and discuss 
what is different for real or imaginary $\mu_{I}$.

\section{QCD INEQUALITY AND PION CONDENSATE}
\label{Sec_QCD_inequality}
We derive QCD inequalities 
for the general meson correlator.
Hereafter,
we impose the condition $m_{u}=m_{d}=m_{0}$.
The meson operator is defined by
\begin{eqnarray}
 M(x)=\bar{q}(x)\Gamma q(x),
\end{eqnarray}
where $\Gamma$ is a product of
the $\gamma$-matrix and the Pauli matrix.
The meson correlator 
then can be written as
\begin{align}
\braket{M(x)M^{\dag}(0)}_{q,A}
 =&-\braket{\textrm{Tr}\left[S(x,0)\Gamma S(0,x)\bar{\Gamma}\right]}_{A}
 \notag \\
 &+\braket{\textrm{Tr}\left[S(x,x)\Gamma\right]}_{A}
 \braket{\textrm{Tr}\left[S(0,0)\bar{\Gamma}\right]}_{A}
 \label{meson_correlator}
\end{align}
with $\bar{\Gamma}=\gamma_{4}\Gamma\gamma_{4}$
\cite{Son-Stephanov, Nussinov_review}.
Here, $\braket{\cdots}_{q,A}$ and $\braket{\cdots}_{A}$
mean
 the full average and the average over the gauge field,
respectively.
The propagator $S(x,y)$ is defined by
$\bra{x}\mathcal{M}^{-1}\ket{y}$
from an inverse fermion matrix. 

Now, we take $\mathcal{\tilde{M}}(i\theta_{q}T,\mu_{I})$
as a fermion matrix, i.e.
the fermion matrix with imaginary $\mu_{q}$
and real $\mu_{I}$.
This matrix satisfies Eq.~(\ref{gamma5_hermiticity_iso})
and hence
Eq.~(\ref{meson_correlator}) can be transformed into
\begin{align}
\braket{M(x)M^{\dag}(0)}_{q,A}
 =&\braket{\textrm{Tr}\left[S(x,0)\Gamma
 \tau^ai\gamma_{5}S^{\dag}(x,0)i\gamma_{5}\tau^a\bar{\Gamma}\right]}_{A}
 \notag \\
 &+\braket{\textrm{Tr}\left[S(x,x)\Gamma\right]}_{A}
 \braket{\textrm{Tr}\left[S(0,0)\bar{\Gamma}\right]}_{A}
 \notag \\
 \le
  &\braket{\textrm{Tr}\left[S(x,0)S^{\dag}(x,0)\right]}_{A}
  \notag \\
  &+\braket{\textrm{Tr}\left[S(x,x)\Gamma\right]}_{A}
 \braket{\textrm{Tr}\left[S(0,0)\bar{\Gamma}\right]}_{A}.
 \label{inequality_iso}
\end{align}
Here, we have employed the Schwartz inequality
to the right-hand side of the first equality.

For imaginary $\mu_{I}$, 
the inequality differs from Eq.~\eqref{inequality_iso}
since the fermion matrix $\tilde{\mathcal{M}}(i\theta_{q}T,i\theta_{I}T)$
satisfies Eq.~(\ref{gamma5_hermiticity_imaginaryiso}),
rather than Eq.~(\ref{gamma5_hermiticity_iso}). 
Adopting the same procedure,
we can obtain
 \begin{align}
  \braket{M(x)M^{\dag}(0)}_{q,A}
 =&\braket{\textrm{Tr}\left[S(x,0)\Gamma
 i\gamma_{5}S^{\dag}(x,0)i\gamma_{5}\bar{\Gamma}\right]}_{A}
 \notag \\
 &+\braket{\textrm{Tr}\left[S(x,x)\Gamma\right]}_{A}
 \braket{\textrm{Tr}\left[S(0,0)\bar{\Gamma}\right]}_{A}
 \notag \\
 \le
  &\braket{\textrm{Tr}\left[S(x,0)S^{\dag}(x,0)\right]}_{A}
  \notag \\
  &+\braket{\textrm{Tr}\left[S(x,x)\Gamma\right]}_{A}
 \braket{\textrm{Tr}\left[S(0,0)\bar{\Gamma}\right]}_{A}
 \label{inequality_imagiso}
 \end{align}
 in the case of imaginary $\mu_{I}$.

Let us take $\Gamma=i\gamma_{5}\tau^a\ (a=1,2)$ and consider the correlator
of the pions $\pi^a$. Note that the
linear combination of $\pi^1$ and
$\pi^2$ gives the charged-pion channel.
In this case, the contribution of a disconnected piece
vanishes for both real and imaginary $\mu_{I}$, 
because
 \begin{eqnarray}
  \braket{\textrm{Tr}
   \left[
    S(x,x)i\gamma_{5}\tau^a
   \right]}_{A}
   &=\braket{\textrm{Tr}
   \left[
    S(x,x)i\gamma_{5}\tau^a(\tau^3)^2
   \right]}_{A}
   \notag \\
  &=-\braket{\textrm{Tr}
   \left[
    S(x,x)i\gamma_{5}\tau^a
   \right]}_{A}
 \end{eqnarray}
holds for $\pi^a$.
Here, we have used $[S(x,x),\tau^3]=0$.
Therefore,
the inequality~(\ref{inequality_iso})
is saturated for $\pi^a$, while
not for $\pi^3$.
The charged-pion condensate 
can thus come out for real $\mu_{I}$.
On the contrary,
the inequality~(\ref{inequality_imagiso}) is not saturated for $\pi^a$,
and hence at least there is no
charged-pion condensate for imaginary $\mu_{I}$.
These statements suggest that the results in Refs.
\cite{Son-Stephanov, Sakai_JPhys}
still hold even when imaginary $\mu_{q}$ is finite.

 \section{CONDENSATE OF NEUTRAL PION}
 \label{Sec_exp_pi3}
 The discussion mentioned above is not applicable for 
 neutral pion, $\pi^3=\bar{q}i\gamma_{5}\tau^3q$,
 since it is isoscalar meson and
 a disconnected piece does not vanish.
 The inequality~(\ref{inequality_iso})
 is saturated only for $\pi^1$ or $\pi^2$, 
 and hence the $\pi^3$ condensate does not occur
 for real $\mu_{I}$~\cite{Son-Stephanov}.
 To prove that $\pi^3$ does not condense also
 for imaginary $\mu_{I}$,
 we should evaluate the expectation value $\braket{\pi^3}$
 directly.
 In this section, we use the framework of
 twisted mass~\cite{Alpha_Coll, Shindler, Hansen, SV}.

 We first define the QCD Lagrangian with imaginary $\mu_{q}$ and $\mu_{I}$,
 together with the twisted mass:
 \begin{align}
  \mathcal{L}_{\rm twist}
  =&\bar{q}(\gamma_{\mu}D_{\mu}+m_{0}-i\gamma_{5}\tau^3 J_{5})q
  \notag \\
  &-i\theta_{q}T\bar{q}\gamma_{4}q-i\theta_{I}T\bar{q}\gamma_{4}\tau^3q
  +\frac{1}{4g^2}F^a_{\mu\nu}F^a_{\mu\nu},
  \label{QCD_Lagrangian_twist}
 \end{align}
 where
 $J_{5}$ is the real parameter determining a twist angle.
 Note that the term $\bar{q}i\gamma_{5}\tau^3J_{5}q$ does
 not change the $\gamma_{5}$-hermiticity~(\ref{gamma5_hermiticity_imaginaryiso}).
 From a generating functional $Z[J_{5}]$ 
 defined by
 \begin{eqnarray}
  Z[J_{5}]
   =\int\mathcal{D}\bar{q}\mathcal{D}{q}\mathcal{D}A_{\mu}
   \exp\left[-\int d^4x\ \mathcal{L}_{\rm twist}\right],
 \end{eqnarray}
 the expectation value of $\pi^3$ is calculated by
  \begin{eqnarray}
  \braket{\pi^3}
   =\lim_{J_{5}\rightarrow 0}\lim_{V\rightarrow\infty}
   \frac{\delta}{\delta J_{5}}
   \log Z[J_{5}],
  \end{eqnarray}
  where $V$ is a volume.

 Now, we first take $m_{0}\ne 0$ and
 rewrite the mass term in Eq.~(\ref{QCD_Lagrangian_twist}) as
\begin{align}
 m_{0}-i\gamma_{5}\tau^3J_{5}
 &=M(J_{5})\textrm{e}^{-i\alpha\gamma_{5}\tau^3},
\end{align}
where $M(J_{5})=\sqrt{m^2_{0}+J^2_{5}}$ and
$\alpha=\tan^{-1}(J_{5}/m_{0})$ is
the twist angle.
Furthermore, we perform the axial $\textrm{U}(1)_{\textrm{I}_{3}}$
transformation to the quark field:
\begin{eqnarray}
 q\ \rightarrow\ \textrm{e}^{i\phi \gamma_{5}\tau^3}q.
  \label{trans_UI3}
\end{eqnarray}
Here, $\phi$ is a rotational angle.
Under this transformation, 
the twisted mass is changed into
\begin{eqnarray}
 M(J_{5})\textrm{e}^{-i\alpha\gamma_{5}\tau^3}
  \ \rightarrow\ M(J_{5})\textrm{e}^{-i(\alpha-\phi)\gamma_{5}\tau^3},
\end{eqnarray}
while the other terms
and the measure in $Z[J_{5}]$ keep the same form.
If we choose $\phi=\alpha$, the twisted-mass term becomes
$M(J_{5})$~\cite{Hansen, SV},
i.e. no phase factor,
and can evaluate $\braket{\pi^3}$ easily.

The fermion matrix we consider is given by
\begin{align}
 \mathcal{\tilde{M}}(i\theta_{q},i\theta_{I};J_{5})
 &=\gamma_{\mu}D_{\mu}+M(J_{5})
 -i\theta_{q}T\gamma_{4}-i\theta_{I}T\gamma_{4}\tau^3
 \notag \\
 &=\mathcal{D}+M(J_{5}),
 \label{fermion_matrix_twisted-mass}
\end{align}
where
$\mathcal{D}=
\gamma_{\mu}D_{\mu}-i\theta_{q}T\gamma_{4}-i\theta_{I}T\gamma_{4}\tau^3$.
The operator $\mathcal{D}$ is an anti-hermitian and hence
its eigenvalue is pure imaginary.
Note that the eigenvalues of the corresponding operator for real $\mu_{I}$
are not purely imaginary in general,
and thereby the discussions 
presented below cannot be applied.

Since the matrix~(\ref{fermion_matrix_twisted-mass}) is
diagonal in flavor space
and the mass $M(J_{5})$
is isospin symmetric,
we can reach the expression
\begin{eqnarray}
 \braket{\pi^3}=\lim_{J_{5}\rightarrow 0}
  \frac{J_{5}}{M(J_{5})}
  \sum_{f=u,d}\int^{\infty}_{0}d\lambda^{f}_{5}
  \frac{\rho(\lambda^{f}_{5},m_{0})}{i\lambda^{f}_{5}+M(J_{5})},
  \label{exp_pi3}
\end{eqnarray}
where $\rho(\lambda^f_{5},m_{0})$ is a spectral function
and $\lambda^f_{5}$ are its eigen values
for each flavor $f$.
It is thus found for $m_{0}\ne 0$
that the $\pi^3$ condensate does not take place
since Eq.~(\ref{exp_pi3}) vanishes.
For $m_{0}=0$,
\begin{eqnarray}
 \braket{\pi^3}
  =\lim_{J_{5}\rightarrow 0}
  \sum_{f=u,d}\int^{\infty}_{0}
  d\lambda^f_{5}
  \frac{\rho(\lambda^f_{5},0)}{i\lambda^f_{5}+J_{5}}
  \label{Banks-Casher-flavor-dependent}
\end{eqnarray}
is deduced, instead of Eq.~(\ref{exp_pi3}).
This relation is equivalent to the
Banks-Casher relation~\cite{BC}
with flavor dependence and
gives the chiral condensate.
For $\mu_{I}=i\theta_{I}T=0$
and $m_{0}=0$,
Eq.~(\ref{Banks-Casher-flavor-dependent})
returns to isospin symmetric Banks-Casher relation
\cite{SV}.

\section{SUMMARY}
\label{Sec_Summary}
In this paper, we 
have investigated appearance of the pion condensate
for real and imaginary $\mu_{I}$,
 introducing imaginary $\mu_{q}$.
 The fermion matrix with imaginary $\mu_{q}$
has positivity for both the cases of
real and imaginary $\mu_{I}$,
but the $\gamma_{5}$-hermiticity is different from each case.

QCD inequalities for the pion correlator were derived,
and the equalities were found to be saturated
for the charged-pion channels when we consider real $\mu_{I}$.
However, for imaginary $\mu_{I}$,
the inequalities are not saturated for the charged pion,
and hence at least any charged-pion condensate does not occur 
for imaginary $\mu_{I}$.
This indicates that the results in previous works~\cite{Son-Stephanov,
Sakai_JPhys} are also true, even when imaginary $\mu_{q}$
is finite.

Finally, we have 
evaluated the expectation value $\braket{\pi^3}$ 
of the neutral pion directly by using the framework of the twisted mass,
because QCD inequalities are not applicable for this channel.
The inapplicability comes from the fact that the $\pi^3$ channel has 
a disconnected piece.
We have proved that $\braket{\pi^3}=0$ 
under the condition $m_{u}=m_{d}=m_{0}>0$. 
For $m_{0}=0$, the Banks-Casher relation 
was derived with flavor dependence.

\noindent
\begin{acknowledgments}
 We thank M. Ishii for fruitful discussions
 and useful comments.
 J.S. also thanks A. Kasai for
 valuable comments.
 J.S. and H. K. are supported by
 Grant-in-Aid for Scientific Research
 (No. 27-7804 and No. 17K05446) from
 the Japan Society for the Promotion of Science (JSPS).
\end{acknowledgments}


\end{document}